\documentclass[aps,prl,reprint]{revtex4-1}

\usepackage[draft]{todonotes}
\usepackage{xspace}

\newcommand{\figref}[1]{Fig. \ref{#1}}
\newcommand{\wcexclude}[1]{}

\usepackage{lipsum}

\renewcommand{\wcexclude}[1]{#1}

\begin{document}

\newcommand{\PaperTitle}{Phase space sampling and operator confidence with generative adversarial networks}

\newcommand\GAN{generative adversarial network\xspace}
\newcommand\gan{generative adversarial network\xspace}
\newcommand\Gan{Generative adversarial network\xspace}
\newcommand\Gans{Generative adversarial networks\xspace}
\newcommand\GANs{generative adversarial networks\xspace}
\newcommand\gans{generative adversarial networks\xspace}

\newcommand{\sqrtbatchsize}{16\xspace}
\newcommand{\batchsize}{256\xspace}
\newcommand{\twobatchsize}{512\xspace}

\newcommand{\generatorlearningrate}{0.0001\xspace}
\newcommand{\discriminatorlearningrate}{0.00002\xspace}
\title{\PaperTitle}

\author{Kyle Mills}
\email[]{kyle.mills@uoit.net}
\affiliation{Department of Physics, University of Ontario Institute of Technology}

\author{Isaac Tamblyn}
\email[]{isaac.tamblyn@nrc.ca}
\affiliation{Department of Physics, University of Ontario Institute of Technology, University of Ottawa \& National Research Council of Canada}

\date{\today}

\begin{abstract}
\wcexclude{
We demonstrate that a \gan can be trained to produce Ising model configurations in distinct regions of phase space.  In training a \gan, the discriminator neural network becomes very good a discerning examples from the training set and examples from the testing set. We demonstrate that this ability can be used as an ``anomaly detector'', producing estimations of operator values along with a confidence in the prediction. 
}
\end{abstract}

\wcexclude{
\maketitle
}

\section{Introduction}

As machine learned potentials come to be used as replacements for conventional force-fields (such as CHARMM, AMBER, etc.) in large-scale and long-time atomistic simulations, it will be important to ensure that these new ``model-free'' methods provide a level of confidence alongside each of their property predictions.

In the field of self-driving vehicles, it has long been recognized that it is more important to develop algorithms which are constantly aware of their accuracy than models which naively return a prediction irrespective of input. Should a vehicle enter an unfamiliar situation (i.e. one that does not exist within the training set), it is important that it identify the drop in predictive confidence so that the human driver can intervene, or the car can safely come to a stop. Naively continuing to make predictions when an algorithm is making large errors is rarely a good plan, and in some cases would be catastrophic.

A similar problem can occur in a numerical simulation based on a configuration-to-energy or configuration-to-force model. When a user is aware that a model is having difficulty making predictions (perhaps because the system has evolved into a fundamentally new region of configuration space, or is violating a conservation law), it is possible to take action. Such actions could include reducing the integration time step or collecting more training data and producing a new model which includes configurations from the new regime.

Without a confidence metric, there is a risk that a model will become unreliable and begin to return unphysical predictions on new data. For the case of molecular dynamics or a Monte Carlo simulation, these erroneous predictions can result in the system venturing further away from the reference set, exacerbating the problem.  The possibility of unphysical predictions outside of the training regime is not a new one in the field of numerical simulation based on fitting. The issue of ``transferability'' is often discussed in the context of force-fields and pseudo-potential construction.

Model-free machine-learned potentials, however, may return predictions which have errors substantially larger in magnitude than those of a conventional force-field. In a traditional force-field expansion, the internal energy and forces acting on system are generally expressed as a sum over bonded and non-bonded terms. The form of these terms are typically set by simple cases which can either be solved analytically, or where the limits are well known. The CHARMM force-field, for example, contains terms relating to bond angles, dihedral angles, and torsions \cite{Brooks1983}.

The fact that force-fields are usually parameterized in terms of simple physical features such as distance, angle, etc, also make it easier to detect when the simulation is approaching an unfamiliar region of configuration space (bond lengths shrink beyond a threshold, for example).

Any form of numerical fitting procedure will be most reliable when predictions are made within the space of training data (i.e. interpolation). Extrapolation, making predictions for properties which are outside of the manifold where data was collected, will invariably result in higher errors. Force-fields which are based on a simple physical expansion tend to be well behaved when extrapolating, as they implicitly contain information about limiting cases through the choice of expansion. The typical Lennard-Jones (6-12 potential) form used for non-bonded interactions, for example, naturally goes to zero at large distances, and diverges when particles become too close.

While such simple functions have the feature that they require very few training examples to make generally acceptable predictions, the cost of this favourable extrapolation behaviour is that these models typically are much less sophisticated than the underlying physics they are designed to describe, and have an upper bound to their accuracy even within the interpolation region.

Model-free machine learned potentials require substantially more data, and therefore the collection of training data becomes important.  The usual concerns with regard to sampling come into play.

In most cases, Nature is able to sample a thermal distribution of states efficiently, although there are notable exceptions such as the glass transition. \emph{In silico}, sampling the distribution of possible configurations is a very difficult task due to the enormous number of free parameters (e.g. position, spin, charge, etc.) defining a near-infinite configuration space for systems of even a modest number of particles.  This ``curse of dimensionality'', for all but the most trivial systems, precludes directly sampling configuration space at non-zero, finite temperature \cite{Carrasquilla2016}. Traditionally, Markov Chain Monte Carlo sampling methods have been devised to obtain random samples from an underlying distribution, but these algorithms, such as Metropolis-Hastings \cite{Metropolis1953, Hastings1970}, depend on the ability to efficiently evaluate both the energy and property of a microstate ($E_i$) and ($O_i$), which can in many cases be a very costly computation. Furthermore, this calculation must be carried out repeatedly, many more times than the desired number of final configurations.

\subsection{\Gans}

Here we propose the use of a generative adversarial network (GAN) \cite{Goodfellow} to carry out both the sampling of configuration space and as a means of providing a confidence estimate for predicted properties.  Generative models are common approaches to unsupervised machine learning \cite{Chen2016}.  \Gans are typically applied to problems in image processing, such as to image super-resolution \cite{Ledig2016}, image-to-image translation \cite{Zhu2017}, and cross-domain pairings (e.g. pairing shoes with matching handbags) \cite{Kim2017}.  They have recently been applied to solutions of differential equations involving transport phenomena \cite{Farimani2017}.

Generative models are based on the core premise that in order to synthesize example data, an understanding of the relevant features must be somehow present in the model, and the training procedure attempts to learn these relevant features, usually in the form of optimizing a set of coefficients in a latent variable space.  A generative adversarial network is a relatively new unsupervised machine learning technique; it is the combination of two ``players'', working against each other as adversaries.  One player acts as a generator, taking random noise as input and producing examples that fall within a probability distribution.  The other player works as a discriminator and learns to tell the difference between examples coming from the generator and true, ground truth examples.  
These players are trained simultaneously with the generator trying to trick the discriminator, and the discriminator learning how to better tell apart the generator's propositions from the true training examples.  A successfully trained \GAN converges to a state where the generator is so good at producing examples that the discriminator cannot tell the generated examples from the ground truth examples.  The key to the success of this method is that the generator is provided with hints about its failure in the form of the gradients, i.e. both the generator and discriminator perform backpropagation using the gradients from the discriminator network. This allows the generator to learn not only whether or not it succeeded in tricking the discrimnator, but effectively gives it access to the reasoning behind its success or failure.

The discriminator can also be provided with relevant labels for the true examples. This acts to help condition the generative adversarial network \cite{Mirza2014}, stabilizing the notoriously sensitive training process \cite{Salimans2016}, but also enabling the trained discriminator to make observable predictions about new data.  Some refer to a network trained with such provisions as a Conditional Generative Adversarial Network (cGAN) \cite{Farimani2017}.  In the case where the discriminator is asked to also reproduce the labels (i.e. the cost function includes an error associated with the labels), the discriminator's output can be interpreted as a likelihood that its label prediction is correct, since the discriminator gets exceptionally good at identifying real examples from the distribution.  In this way, the discriminator of the trained \gan can be used as an anomaly detector, providing label predictions alongside a probability that the prediction is correct.

In theory, the generator and discriminator could be any learning algorithm capable of backpropagation.  In this work, we use deep convolutional neural networks as their success on the Ising model has been demonstrated in supervised machine learning classification and prediction \cite{Mills2017a,Morningstar2017}.

\subsection{The Ising Model}
The square two-dimensional Ising model is a well-studied example of a ferromagnetic system of particles \cite{Onsager1944}.  The model consists of an $L\times L$ grid of discrete interacting ``particles'' which either possess a spin up ($\sigma = 1$) or spin down ($\sigma = -1$) moment.  The internal energy associated with a given configuration of spins is given by the Hamiltonian $\hat H = -J \sum\sigma_i\sigma_j  $
where the sum is computed over all nearest-neighbour pairs ($\langle i,j\rangle$), and $J$ is the interaction strength. For $J=1$, the system behaves ferromagnetically; there is an energetic cost of having opposing neighbouring spins, and neighbouring aligned spins are energetically favourable.  A measure of ``disorder'' in the system can be represented by an order parameter known as the ``magnetization'' $M$, which is merely the average of all $L^2$ individual spins.  Because both the internal energy and magnetization depend on the discrete spins within the system, both quantities exhibit a discrete distribution.  The configuration space of the $8\times 8$ Ising model is of size $2^{8^2}$, and thus sampling from all possible configurations is impossible. Given a configuration, while computing properties of interest (e.g. energy) is generally possible through some theoretical framework, the reverse is not true; it is not possible to obtain a configuration that satisfies a given property.
In order to generate examples of a specific energy, one must employ some form of Monte Carlo approach, or devise another clever sampling algorithm \cite{Portman2016}.  Because of its simplicity and ubiquity, the Ising model has made many recent appearances in machine-learning investigations \cite{Carrasquilla2016,Mills2017a,Luchak2017,Wang2016,Tanaka2016}

\section{Methods}

\subsection{Training datasets}

We used a targeted sampling procedure \cite{Mills2017a} resembling the Metropolis-Hastings algorithm to produce four datasets used for training:  
\begin{itemize}
\item A low-energy dataset, generated by using a target energy of $-1.3L^2$,
\item a high-energy dataset, generated by using a target energy of $+1.3L^2$,
\item a bimodal dataset consisting of a equal combination of the previous two datasets, and
\item a dataset consisting of an equal number of configurations from each of the 63 possible energy values.
\end{itemize}
The energy distributions of the first three datasets are displayed as the dashed lines in \figref{distributions}.

\subsection{Network}

\begin{figure}
\begin{center}
 \includegraphics[width=0.75\columnwidth]{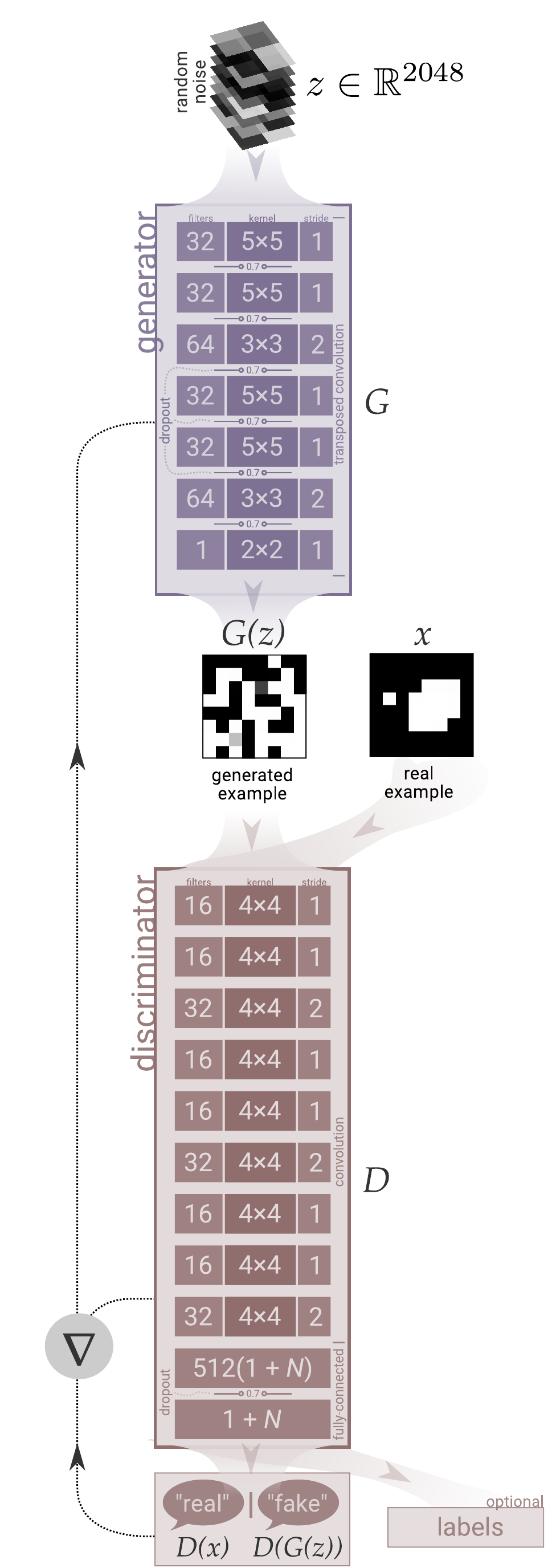}
 \caption{The \GAN architecture we used.  The generator takes random noise, and through a series of transposed convolutions, produces an example.  The discriminator takes true examples and ``fake'' examples from the generator and predicts the probability that each is a real example.  The gradients from the discriminator are used by both adversaries to improve their weights. \label{schematic}}
 \end{center}
\end{figure}

The generator takes an array of random noise $z$ as input and produces examples attempting to mimic the true distribution.  Our generator function takes a $2\times 2\times 128$ random array sampled from a zero-centered normal distribution with standard deviation of 0.7.  It passes the random data through 7 transposed convolution layers with varying kernel, stride, and filter counts arriving at the appropriately shaped $8\times 8$ output, $G(z)$.  After every transposed convolutional layer (with the exception of the last two), we include a dropout layer with a 0.7 retention probability, to hopefully prevent the generator from memorizing the training set.  Since individual spins of the Ising model can either be $1$ or $-1$, all transposed convolution layers have hyperbolic tangent (tanh) activation, which produces output at an appropriate scale.

The discriminator takes an $8\times 8$ array as input.  This can either be $G(z)$ from the generator, or an example $x$ from the training set. Through a series of 9 convolutional layers, and two fully connected layers it reduces the example to a single output $D$.  In the case where we provide labels to the discriminator, the output is of size $1 + N$, where $N$ is the number of conditioning labels provided to the discriminator.  All layers except for the final layer use rectified linear unit (ReLU) activation.  We use a dropout layer with a 0.7 retention probability between the final two fully-connected layers. 

We implemented the model in TensorFlow \cite{GoogleResearch2015}.  If trained naively, we discovered that the discriminator learned much more quickly than then generator.  When this happened, the discriminator did not provide sufficient feedback (gradients) to the generator and therefore the generator was unable to improve.  To remedy this problem, we used a learning rate five times greater for the generator, essentially handicapping the discriminator, and leading to better convergence of both opponents.

The objective of the generator is to minimize the loss function
\begin{equation}
L^{\mathrm{(G)}} = -\log(D(G(z)),
\end{equation}
and the objective of the discriminator is to minimize the loss function
\begin{equation}
L^{\mathrm{(D)}}_{\mathrm{GAN}} = \log(1-D(G(z))) + \log(D(x)),
\end{equation}
while simultaneously minimizing the mean-squared error of the labels $y$:
\begin{equation}
L^{\mathrm{(D)}}_{\mathrm{LABEL}} = \frac{1}{N_\mathrm{E}} \sum\limits_{i=0}^{N_\mathrm{E}} \sum\limits_{j=0}^{N_\mathrm{L}}\alpha_j\left(y_j^{\mathrm{(predicted)}} - y_j^{\mathrm{(true)}}\right)^2
\end{equation}
where $N_\mathrm{E}$ is the number of training examples, $N_\mathrm{L}$ is the number of conditioning labels (in our case $N_\mathrm{L}=2$: energy and magnetization), and $\alpha_j$ is an optional label scaling parameter to account for the relative importance and/or scale when using multiple conditioning labels (for example, the range of Ising energies is always twice that of the Ising magnetizations, so correcly predicting energies would be considered ``more important'' by the discriminator unless scaled appropriately). The discriminator loss is simply a weighted sum of these two individual losses:
\begin{equation}
L_D = \beta L^{\mathrm{(D)}}_{\mathrm{GAN}} + \gamma L^{\mathrm{(D)}}_{\mathrm{LABEL}}.
\end{equation}

We found we obtained the best performance when we initially set $\beta=\gamma=0.5$.  After training for 1000 epochs, we reduced $\gamma$ to $0.02$ and left $\beta$ unchanged. This permitted the discriminator, which had at this point learned to accurately predict the labels, to ``focus'' on its ability to differentiate between real and fake outputs, improving anomaly detection.

Contrary to supervised learning, monitoring the loss functions is not an informative method to verify convergence of the model.  Both players are simultaneously trying to reduce their own loss functions; a decrease in one leads to an increase in the other.  Therefore, during training, we periodically plot the energy and magnetization distributions produced by the generator to verify they are approaching the training distributions.

\begin{figure}
 \includegraphics[width=0.999\columnwidth]{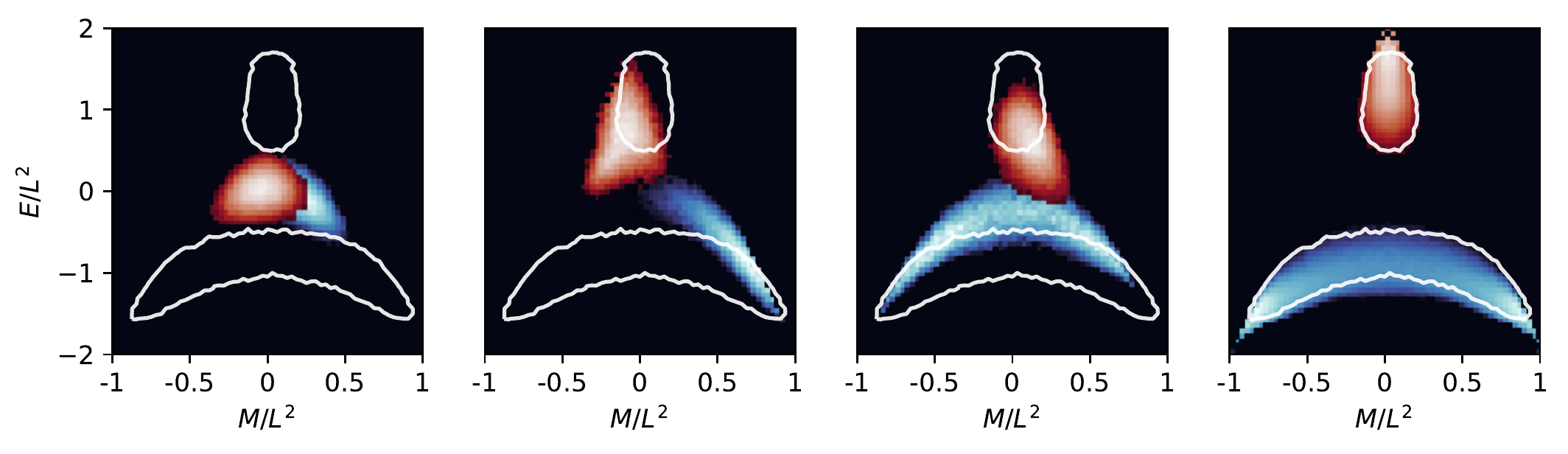}
 \caption{An example progression of the generator through phase space during the training process for the high energy and low energy target distributions.  The generated distributions are represented by the heatmap, and the white boundary represents the target (i.e. training) distribution. \label{phase_space_progression}}
\end{figure}

\begin{figure}
 \includegraphics[width=0.999\columnwidth]{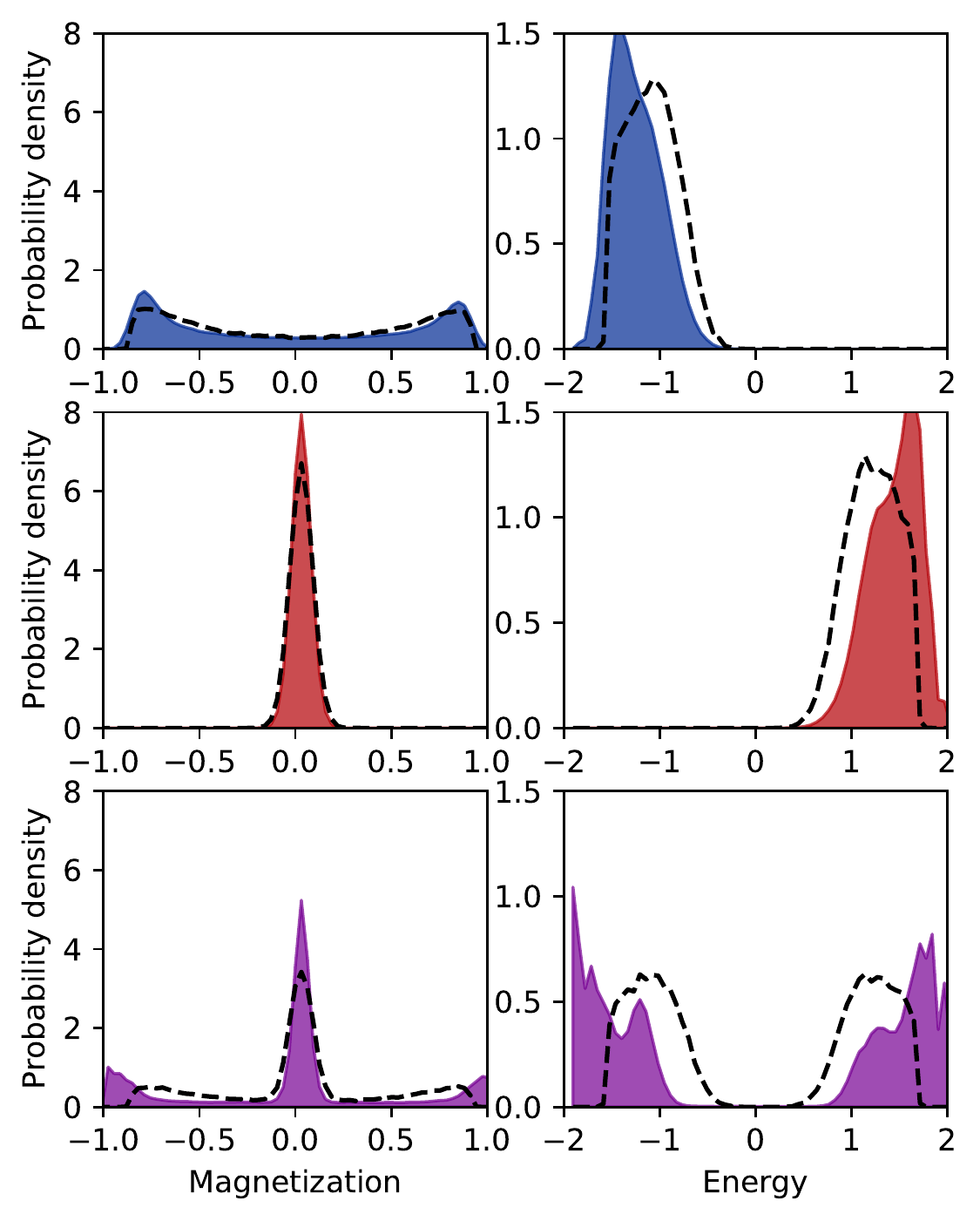}
 \caption{The energy and magnetization distributions for the training data (dashed lines) and generator output (solid fill).  We do not specifically request that the \GAN reproduce these distributions; it  must do so automatically.  While not matching the exact shape, the \GAN is able to produce examples from the same energy and magnetization ranges, permitting efficient sampling.   \label{distributions}}
\end{figure}

Training proceeds by passing a batch of \twobatchsize images to the discriminator.  \batchsize images are from the training set, and \batchsize are the output from the generator. Initially, the generator has not learned how to produce realistic looking examples, and outputs nothing more than random noise.  We use the Adam optimization method \cite{Kingma2014} with a learning rate of \discriminatorlearningrate to update the discriminator's weights.  Then it is the generator's turn; we feed a batch of random arrays to the generator.  The generator uses its learned filters to process the random information into realistic-looking Ising configurations.  Of course, the generator performs very poorly during the first few iterations, but the generator's Adam optimizer receives feedback from the discriminator and is able to improve on its kernels to produce better examples.  This process is repeated thousands of times, until the energy and magnetization distributions (Figs. \ref{phase_space_progression} and \ref{distributions}) are deemed sufficiently similar.  We used a total of \mbox{$N_{\mathrm{E}} = 50,000$} training images for each \gan we trained.

\section{Results}

We trained three \GANs: one on the low-energy distribution, one on the high-energy distribution, and one on the bimodal mixture of high- and low-energy distributions.  Shortly after training begins, the generator produces examples composed of random spins, and therefore the distributions of energy and magnetization are mostly centered at the zero-point (\figref{phase_space_progression}).  As the generator and discriminator learn from each other, the generator begins to better match the training distributions.  In the final frame of \figref{phase_space_progression}, the high-energy distribution (red) matches almost exactly with the high-energy training distribution (white outline), and the low-energy distribution (blue) matches closely with its corresponding training distribution.  The third row in \figref{distributions} shows the performance of the bimodal dataset. The trained \gan shown here was one of the few training runs that, with moderate success, captured both modes of the distribution.  In many training runs, the generator collapsed to either the high-energy or low-energy mode.  This ``mode-collapse'' in \gans is a common, and understood phenomena with ongoing research investigating possible solutions \cite{Srivastava2017,Metz2016,Tolstikhin2017,Salimans2016a}.

In training the \gan, the discriminator by design becomes very good at identifying configurations which fall outside of the desired distribution, with its output falling roughly on a scale between -1 (the example is suspected to be ``fake'', i.e. not from the training distribution) and 1 (the example is deemed ``real'', i.e. likely to be from the training distribution).  Provided the discriminator is also able to produce an estimate of the value of an operator (e.g. magnetization and/or energy labels were provided to the discriminator during training), then the output of the discriminator can be interpreted as a confidence of its operator prediction.  Thus the \gan process produces an anomaly detector, providing both a continuous, real-valued operator evaluation (regression) as well as an indication that the predicted value is correct.

\begin{figure}
 \includegraphics[width=0.999\columnwidth]{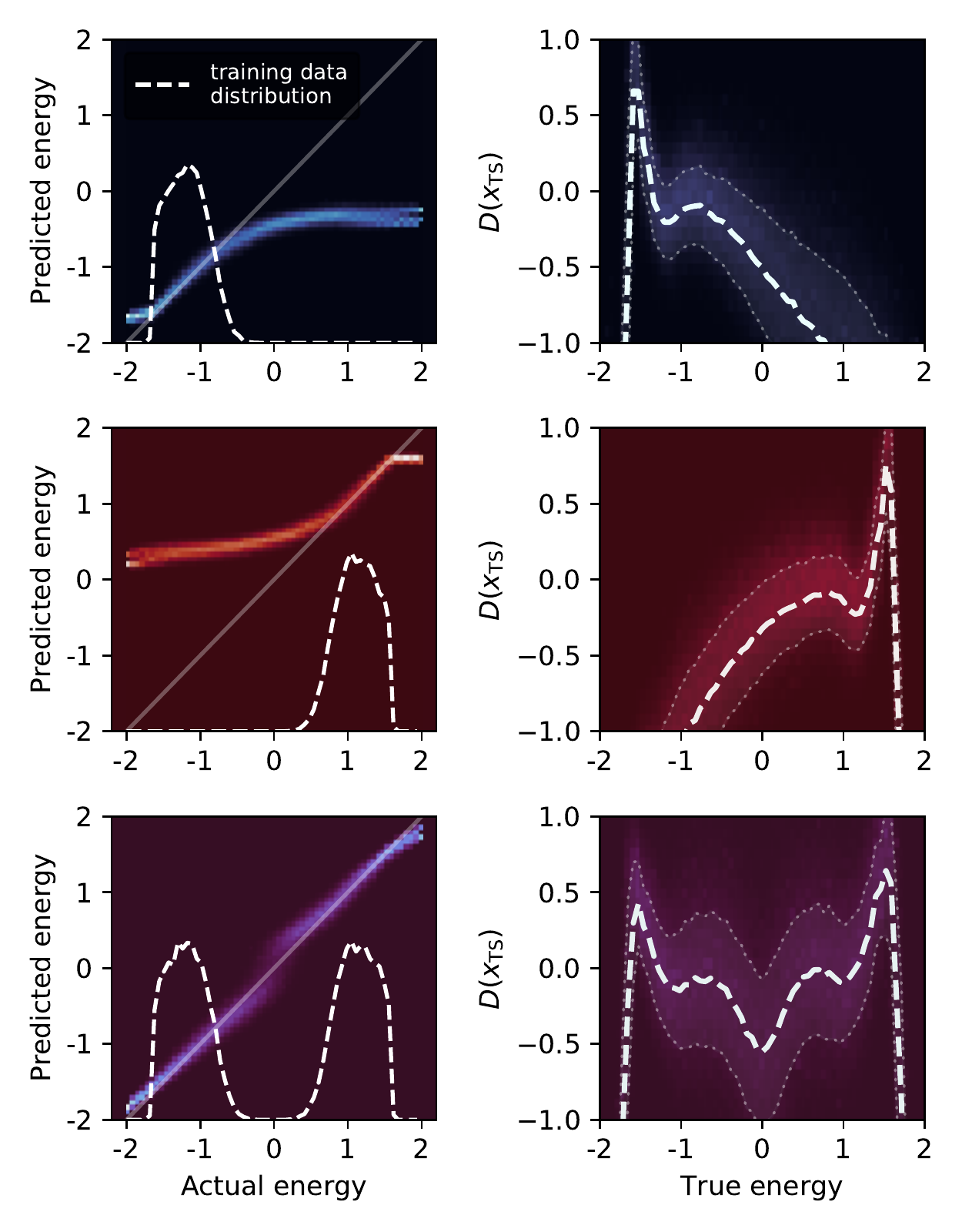}
 \caption{In addition to its task of opposing the generator, the discriminator can learn to make predictions about operator values if provided with the corresponding labels.  Left column: histograms of the predicted vs. true energy for examples from the training set.  Right column: when trained using a label, the discriminator's output can be interpreted as a measure of confidence in its prediction. When provided with a uniform distribution of energies across the entire energy range, the discriminator has a much higher average confidence in the region it has not yet seen.\label{anomaly}}
\end{figure}

After training the discriminator, we presented it with examples drawn from an even distribution across the energy range.  In such an application, one would expect that the discriminator makes predictions both \textit{accurately and confidently} near the region on which it was trained. Arguably more importantly is the ability of the discriminator to output a low confidence when it is unsure of the answer, such as in the regions where training data was not provided. \figref{anomaly} shows that this is indeed the case; the discriminator outputs the highest confidence in regions where it was provided training examples.  In regions devoid of training examples, the discriminator indicates its incompetence in making predictions by outputting a low confidence, precisely what one would desire of a practical prediction engine.

\section{Conclusion}

We have trained a \gan to be able to efficiently sample phase space, producing examples from a target distribution without the necessity of \textit{a priori} knowledge of the distribution.  A \gan uses to separate convolutional neural networks, the generator and the discriminator.  The generator is tasked with producing examples so realistic that the discriminator cannot tell them apart, and the two networks are trained in tandem.  Ultimately, the generator becomes so good at producing realistic examples that the discriminator cannot differentiate between the training examples and the output of the generator.  If one provides labelled data to the discriminator, we show that the discriminator can be trained to make accurate predictions as well as indicate its confidence in the predictions, essentially anomaly detection through supervised learning.

The generator of a trained \gan can be used to produce examples which are larger than those on which it was trained \cite{Bergmann2017}. This has been used to produce textures which have arbitrarily large spatial extent, but are based on a spatially-finite training set.  Successful application of this technique to physical systems would be incredibly valuable.  As the spatial extent of a physical system increases, the individual features comprising the system do not increase in size, but rather in number (extensivity).  The simulation cost for such a system, however, increases dramatically.  Therefore, a \gan trained on a small system which is capable of producing examples from a spatially larger distribution would be incredibly useful, and we propose this as a future application of \gans.

\Gans are notoriously difficult to train, and the training process can be quite unstable, however ongoing research aimed at stabilizing the training process and making \gans more robust will lead to \gans as a promising tool for use in the physical sciences.

\wcexclude{
\section{Acknowledgements}
The authors acknowledge funding from NSERC and SOSCIP. Compute resources were provided
by the National Research Council of Canada, SOSCIP, and Compute Canada. 
}

\wcexclude{
\bibliography{main_flat}
}

\clearpage

\end{document}